\documentclass[12pt]{article}

\usepackage{a4wide}
\usepackage[pdftex,usenames,dvipsnames]{color}
\usepackage{graphics}
\usepackage{amsfonts}
\usepackage{amssymb}

\newcommand{\nit}{\noindent}

\newcommand{\np}{\newpage}
\newcommand{\dsp}{\displaystyle}
\newcommand{\vs}[1]{\vspace{#1 ex}}
\newcommand{\hs}[1]{\hspace{#1 em}}
\newcommand{\bfr}{\begin{flushright}}
\newcommand{\efr}{\end{flushright}}
\newcommand{\bc}{\begin{center}}
\newcommand{\ec}{\end{center}}
\newcommand{\ben}{\begin{enumerate}}
\newcommand{\een}{\end{enumerate}}

\newcommand{\be}{\begin{equation}}
\newcommand{\ee}{\end{equation}}
\newcommand{\ba}{\begin{array}}
\newcommand{\ea}{\end{array}}
\newcommand{\ct}{\cite}
\newcommand{\bit}{\bibitem}

\newcommand{\eps}{\epsilon}

\newcommand{\lb}{\lambda}

\newcommand{\rg}{\rho}

\newcommand{\vf}{\varphi}
\newcommand{\og}{\omega}

\newcommand{\Del}{\Delta}

\newcommand{\lh}{\left(}
\newcommand{\rh}{\right)}

\begin{document}

\pagestyle{empty} 

\begin{center} 
{\Large{\bf Single scalar cosmology}} \\
\vs{7} 

{\large J.W.\ van Holten}\\ 
\vs{5}

{\large NIKHEF}
\vs{2} 

{\large Amsterdam NL} \\
\vs{2}

and \\
\vs{2}

{\large Instituut Lorentz, Leiden University}\\
\vs{2}

{\large Leiden NL}\\
\vs{7}

\today
\end{center} 
\vs{3}

\nit
{\small
{\bf Abstract} \\
The cosmology of flat FLRW universes dominated by a single scalar field is discussed. General features of the 
evolution of the universe and the scalar field are illustrated by specific examples. It is shown that in some situations
the most important contribution to inflation comes from the approach to a region of slow roll, rather than from  
the period of leaving a slow-roll regime.
}
 
\np
~\hfill

\np

\pagestyle{plain}
\pagenumbering{arabic} 

\section{Scalar fields in cosmology}

The discovery of the Higgs particle \ct{atlas2012,cms2012} has confirmed that scalar fields play a fundamental 
role in subatomic physics. Therefore they must also have been present in the early universe and played a part 
in its development \ct{veltman1974,guth1981,steinhardt-albrecht1982,linde1983,turner1983}. Significant 
evidence is provided by observations of the CMB \ct{cobe1994,wmap2012,planck2013}. 
About scalar fields on present cosmological scales nothing is known, but in view of the observational evidence 
for accelerated expansion \ct{riess_etal1998, perlmutter_etal1999} it is quite well possible that they take part 
in shaping our universe now and in the future \ct{wetterich1988,zlatev_etal1999}. 

Understanding the impact of scalar fields on the evolution of the cosmos is therefore of direct observational 
relevance. There is already a vast literature on the subject, e.g.\ \ct{copeland_etal1993}-\ct{vholten2013} and 
references therein. In this paper I confine myself to a summary of some recent work concerning the simplest 
scenario, the evolution of a flat, isotropic and homogeneous universe in the presence of a single cosmic 
scalar field. Neglecting ordinary matter and radiation, the evolution of such a universe is described by two 
degrees of freedom, the homogeneous scalar field $\vf(t)$ and the scale factor of the universe $a(t)$. The 
relevant evolution equations are the Friedmann and Klein-Gordon equations, 
reading\footnote{We use units in which $c = \hbar = 8 \pi G = 1$.}
\be
\frac{1}{2}\, \dot{\vf}^2 + V = 3 H^2, \hs{2} \ddot{\vf} + 3 H \dot{\vf} + V' = 0,
\label{1.1}
\ee
where $V[\vf]$ is the potential of the scalar fields, and $H = \dot{a}/a$ is the Hubble parameter. 
Furthermore, an overdot denotes a derivative w.r.t.\ time, whilst a prime denotes a derivative w.r.t.\ the
scalar field $\vf$. 

For single-field models in which the scalar $\vf(t)$ is a single-valued function in some interval of time, 
it is possible to reparametrize the Hubble parameter in terms of $\vf$:
\be
H(t) = H[\vf(t)].
\label{1.2}
\ee
This will be assumed in the following. Now taking time derivatives, we arrive at the results
\be
\dot{\vf} \lh \ddot{\vf} + V' \rh = 6 H \dot{H}, \hs{2} \dot{H} = H' \dot{\vf}.
\label{1.3}
\ee
It follows, that for $\dot{\vf} \neq 0$ and $H \neq 0$ one gets
\be
\dot{\vf} = - 2 H', \hs{2} \dot{H} = - \frac{1}{2}\, \dot{\vf}^2 \leq 0.
\label{1.4}
\ee
Thus the Hubble parameter is a semi-monotonically decreasing function of time. 
Finally, replacing the time derivatives in the Friedmann equation we find 
\be
V = 3 H^2 - 2 H^{\prime\, 2}.
\label{1.5}
\ee
Given $V[\vf]$ this is a first-order differential equation for $H[\vf]$. In the following we discuss the solutions 
of the is equation. 

\section{Stationary points \label{s2}} 

As we have observed, $H(t)$ is a decreasing function of time, except at stationary points where
$\dot{\vf} = - 2H' = 0$. It follows first of all, that as long as $H$ is positive, the expansion of the universe 
will generically slow down, whilst in case $H(t)$ crosses into the domain of negative values collapse of 
the universe becomes inevitable. Therefore the existence or non-existence of stationary points is relevant 
to the question of  the ultimate fate of our simple model universes. 

There are two kinds of stationary points; a point where $\dot{\vf} = H' = 0$ is an end point of the evolution if
\be
\ddot{\vf} = 4 H' H'' = 0,
\label{2.1}
\ee
which happens if $H''$ is finite. In contrast, if 
\be
\ddot{\vf} = 4 H' H'' \neq 0,
\label{2.2}
\ee
$H''$ necessarily diverges in such a way as to make $\ddot{\vf}$ finite: $H'' \propto 1/H'$.
This property is illustrated by the remarkable example of the eternally oscillating scalar field:
\be
\vf(t) = \vf_0 \cos \og t.
\label{2.3}
\ee
For such a scalar field to exist we require
\be
H' = - \frac{1}{2}\, \dot{\vf} = \frac{\og \vf_0}{2} \sin \og t = \frac{\og}{2} \sqrt{\vf_0^2 - \vf^2}.
\label{2.4}
\ee
It is clear, that there are infinitely many stationary points 
\be 
\og t_n = n \pi, \hs{2} \vf(t_n) = (-1)^n \vf_0,
\label{2.5}
\ee
where $H' = 0$. Now
\be
H'' = - \frac{1}{2} \frac{\og \vf}{\sqrt{\vf_0^2 - \vf^2}},
\label{2.6}
\ee
and therefore $H''$ diverges at all stationary points $t_n$, but  in such a way that 
\be
4 H' H'' = - \og^2 \vf = \ddot{\vf}.
\label{2.6.1}
\ee
We conclude that all stationary points (\ref{2.5})  are turning points.

However, though the scalar field oscillates forever with constant frequency and amplitude, 
this does not imply that the scale factor oscillates as well; on the contrary, the Hubble 
parameter ultimately moves into the domain of negative values and the universe collapses:
\be
\ba{lll}
H & = & \dsp{ 
 H_0 - \frac{1}{4} \og \vf_0^2 \arccos \lh \frac{\vf}{\vf_0} \rh +\frac{1}{4}\, \og \vf \sqrt{\vf_0^2 - \vf^2} }\\
 & & \\
 & = & \dsp{ H_0 - \frac{1}{4}\, \og^2 \vf_0^2\, t + \frac{1}{8}\, \og \vf_0^2 \sin 2 \og t.}
\ea
\label{2.7}
\ee
The corresponding solution for the scale factor is
\be
a(t) = a(0) e^{H_0 t - \frac{1}{8} \og^2 \vf_0^2 t^2 + \frac{1}{16} \lh 1 - \cos 2 \og t \rh},
\label{2.8}
\ee
which is a gaussian, slightly modulated by an oscillating function of time; see fig.\ 1.

\bc
\scalebox{0.3}{\includegraphics{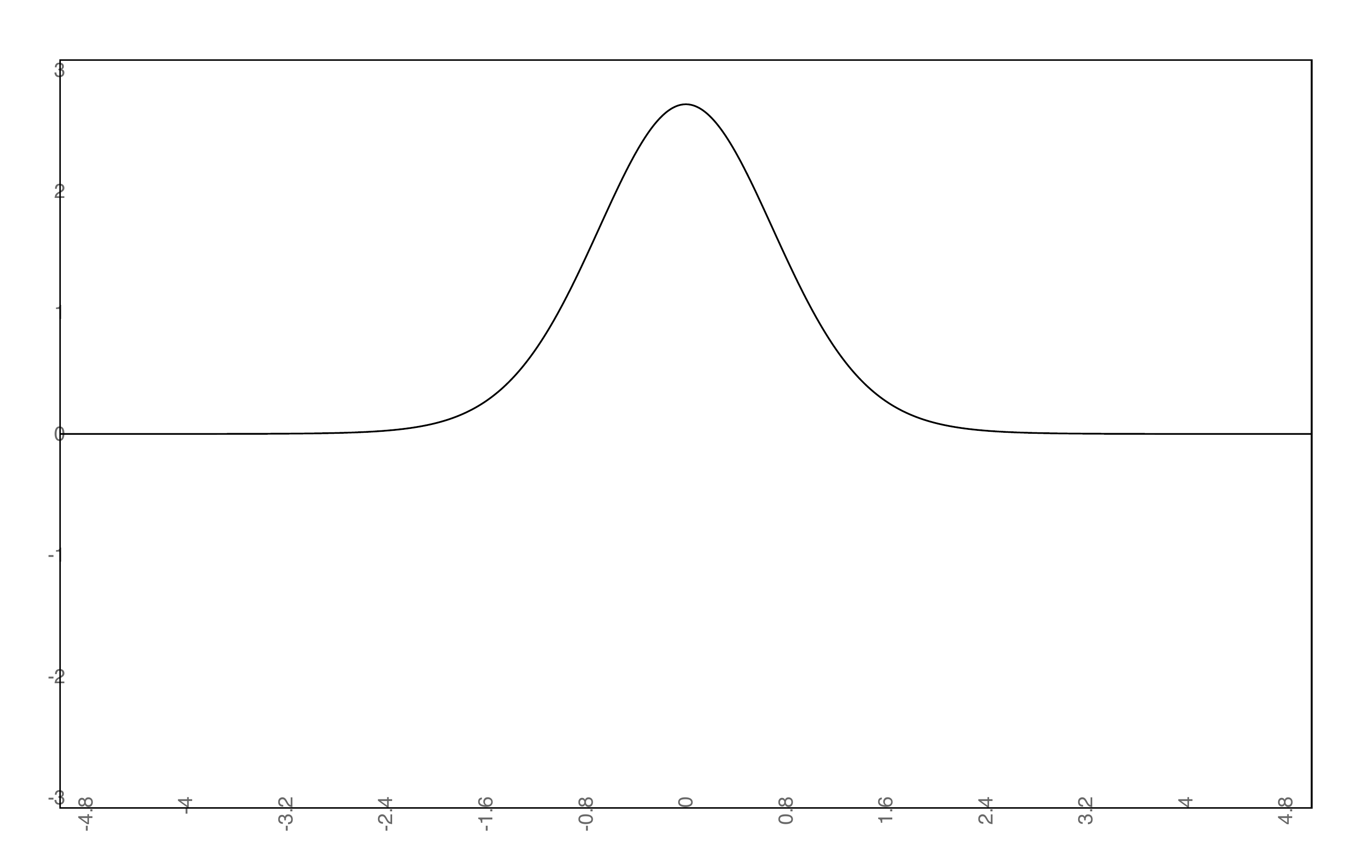}}
\vs{1}

{\footnotesize{Fig.\ 1: Scalefactor $a(t)$ for an eternally oscillating scalar field.}}
\ec

\vs{-25} \scalebox{0.8}{\hs{7} a(t)} \\
               \scalebox{0.8}{\hs{11} 0}
\vs{25}

\nit
The potential giving rise to this behaviour can also be constructed:
\be
\ba{lll}
V & = & \dsp{  3 H^2 - 2 H^{\prime\, 2} }\\
 & & \\
 & = & \dsp{ 3 \lh H_0 - \frac{1}{4}\, \og \vf_0^2 \arccos \lh \frac{\vf}{\vf_0} \rh + \frac{1}{4} \og \vf 
  \sqrt{ \vf_0^2 - \vf^2} \rh^2 - \frac{\og^2}{2} \lh \vf_0^2 - \vf^2 \rh. }
\ea
\label{2.9}
\ee
Observe, that this potential keeps track of the number of oscillations the scalar field has performed through 
the arccos-function, so ultimately $V$ increases indefinitely as a function of time, whilst the volume of a 
representative domain of space decreases rapidly. 

\section{The fate of the universe}

The arguments presented above bring out that there are only a few final states for universes governed by 
a single scalar field at large times. Once $H$ becomes negative the collapse of the universe becomes 
inevitable; if $H$ never becomes negative, it still must tend to a vanishing or positive final value, which 
can be reached either in finite or infinite time. The universe then ends up in a Minkowski or in a de Sitter 
state. These conclusions are a consequence of the non-positivity of $\dot{H}$, eq.\ (\ref{1.4}), which 
implies that a negative $H$ can never return to larger values at later times 
\ct{faraoni-protheroe2012,vholten2013}. 

In order to establish the existence of end points or asymptotic end points of evolution at non-negative 
values of $H$, we first consider the locus of all stationary points, defined by 
\be
\dot{\vf} = - 2 H' = 0 \hs{2} \Rightarrow \hs{2} V = 3 H^2 \geq 0 .
\label{3.1}
\ee
It follows that stationary points can occur only in the region of positive or vanishing potential. 
In particular this holds for end points, which therefore do not occur in a region of negative potential. 
Moreover, it is clear that a Minkowski final state occurs only at a stationary point where $V = 0$, whereas  
all stationary points with $V >  0$ correspond to de Sitter states. To correspond to an end point of the 
evolution, $H''$ must be finite at these stationary points to guarantee that $\ddot{\vf} = 0$ as well.

Next observe, that eq.\ (\ref{1.5}) implies
\be
V' = 2 H' (3 H - 2 H''),
\label{3.2}
\ee
and therefore $V' = 0$ if $H' = 0$ and $H$ and $H''$ are finite. As a result end points of 
the evolution necessarily occur at an extremum of $V$, but only if $V \geq 0$ there.

We can illustrate these results by the example of quadratic potentials 
\be
V = v_0 + \frac{m^2}{2}\, \vf^2.
\label{3.3}
\ee
We distinguish the cases $v_0 > 0$, $v_0 = 0$ and $v_0 < 0$. The stationary points (\ref{3.1})  
are represented graphically by the curves in the $\vf$-$H$-plane in fig.\ 2.

\bc
\scalebox{0.19}{\hs{-4}\includegraphics{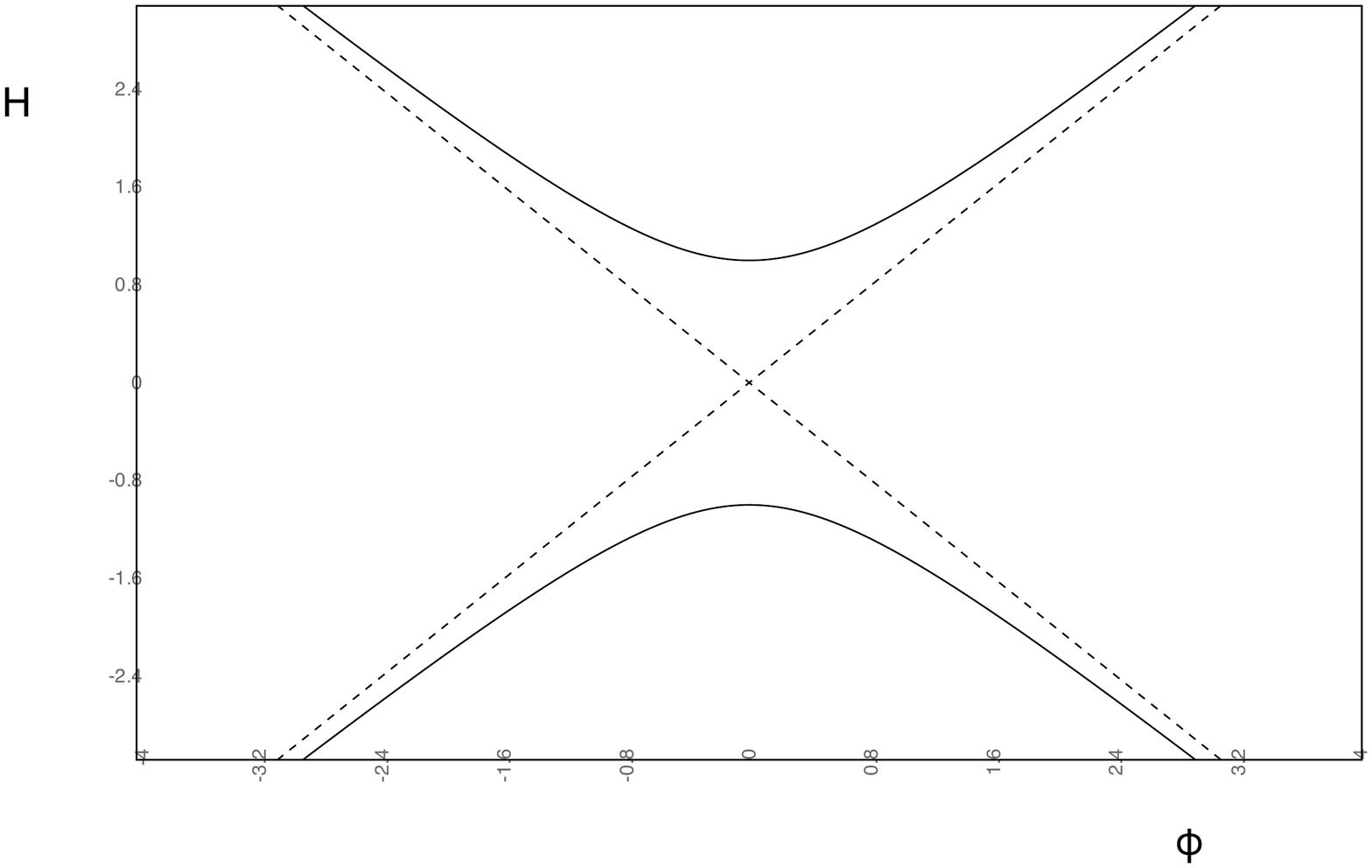}, \includegraphics{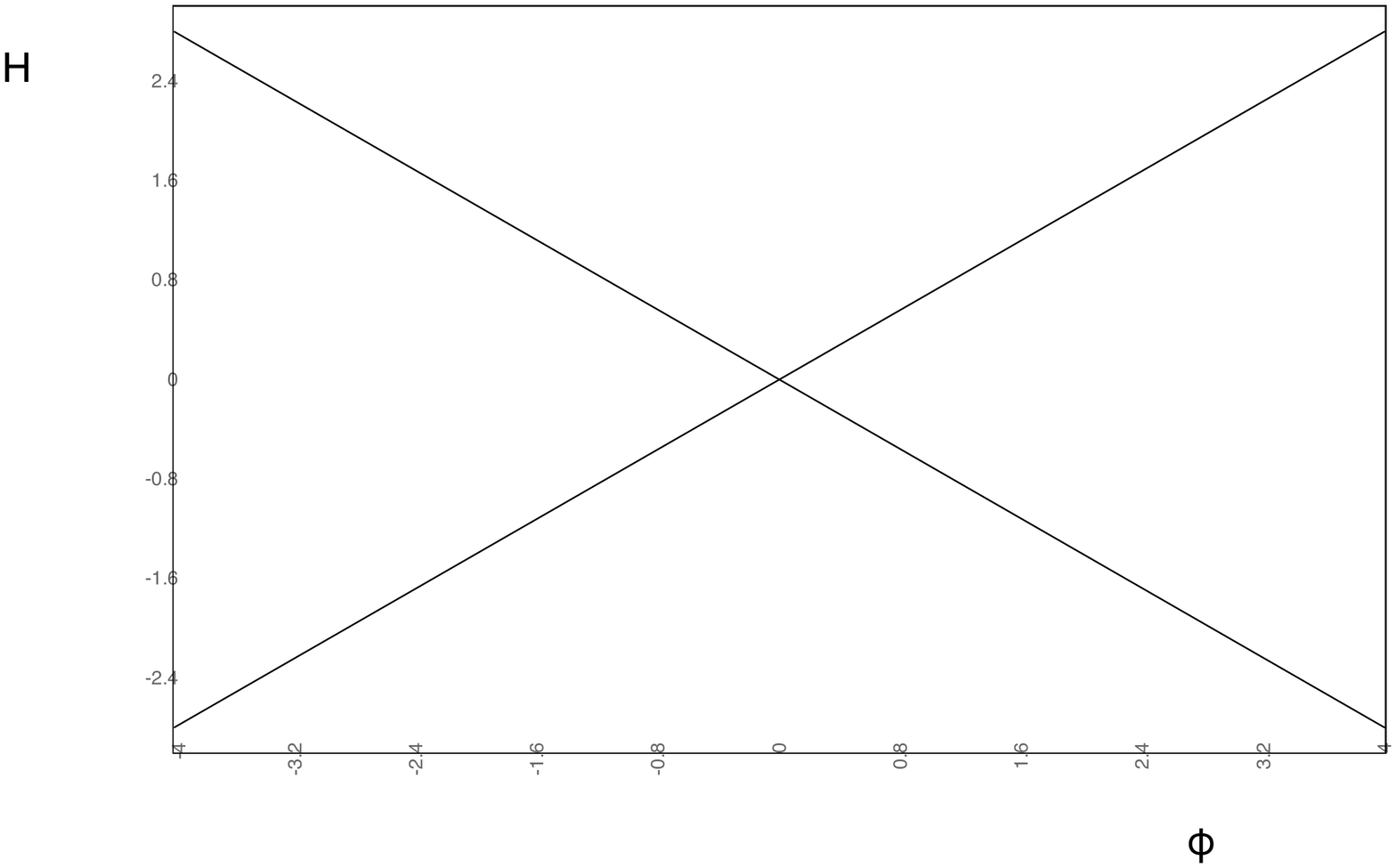}, \includegraphics{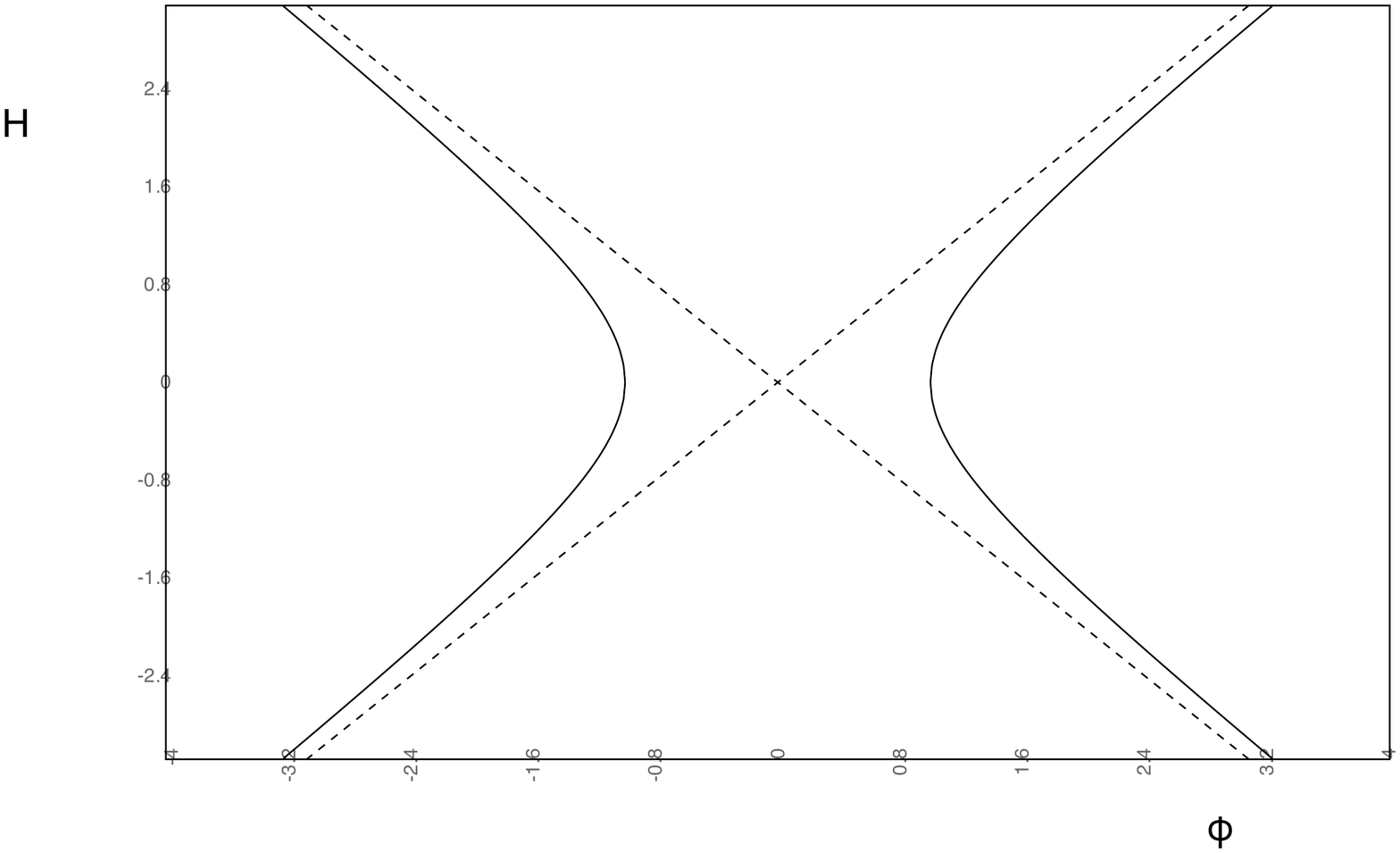}}

{\footnotesize{\hs{1} Fig.\ 2: Critical curves $H'[\vf] = 0$ for quadratic potentials (\ref{3.3}) with $v_0 > 0$ (left), 
$v_0 = 0$ (middle) and $v_0 < 0$ (right).}}
\ec

\nit
First, for $v_0 > 0$ there exists a stationary point for any value of $\vf$, but the potential has a unique 
minimum at $\vf = 0$, which is the only stationary point where $V' = 0$, and therefore the only end point. 
Indeed, once this point is reached $H$ can not decrease anymore and we have final state of de 
Sitter type. For $v_0 = 0$ the critical curves become straight lines, crossing at the origin where $H = 0$ at 
$V = 0$. This is still a stationary point with $\ddot{\vf} = 0$ representing a Minkoswki state, but as $V'$ is not
defined there it is really to be interpreted as a limit of the previous case. There are no evolution curves flowing
from the domain $H > 0$ to the domain $H < 0$. Finally, for $v_0 < 0$ there are no stationary points in the 
region $\vf^2 < 2 |v_0| /m^2$, and the solutions can cross into the domain of negative $H$ there. 

Fig.\ 3 shows some actual solutions for the cases of $v_0 > 0$ and $v_0 < 0$, respectively. These solutions
can be obtained by making a power series expansion \ct{vholten2013} 
\be
H[\vf] = h_0 + h_1 \vf + h_2 \vf^2 + h_3 \vf^3 + ...
\label{3.4}
\ee

\bc
\scalebox{0.17}{\hs{-3}\includegraphics{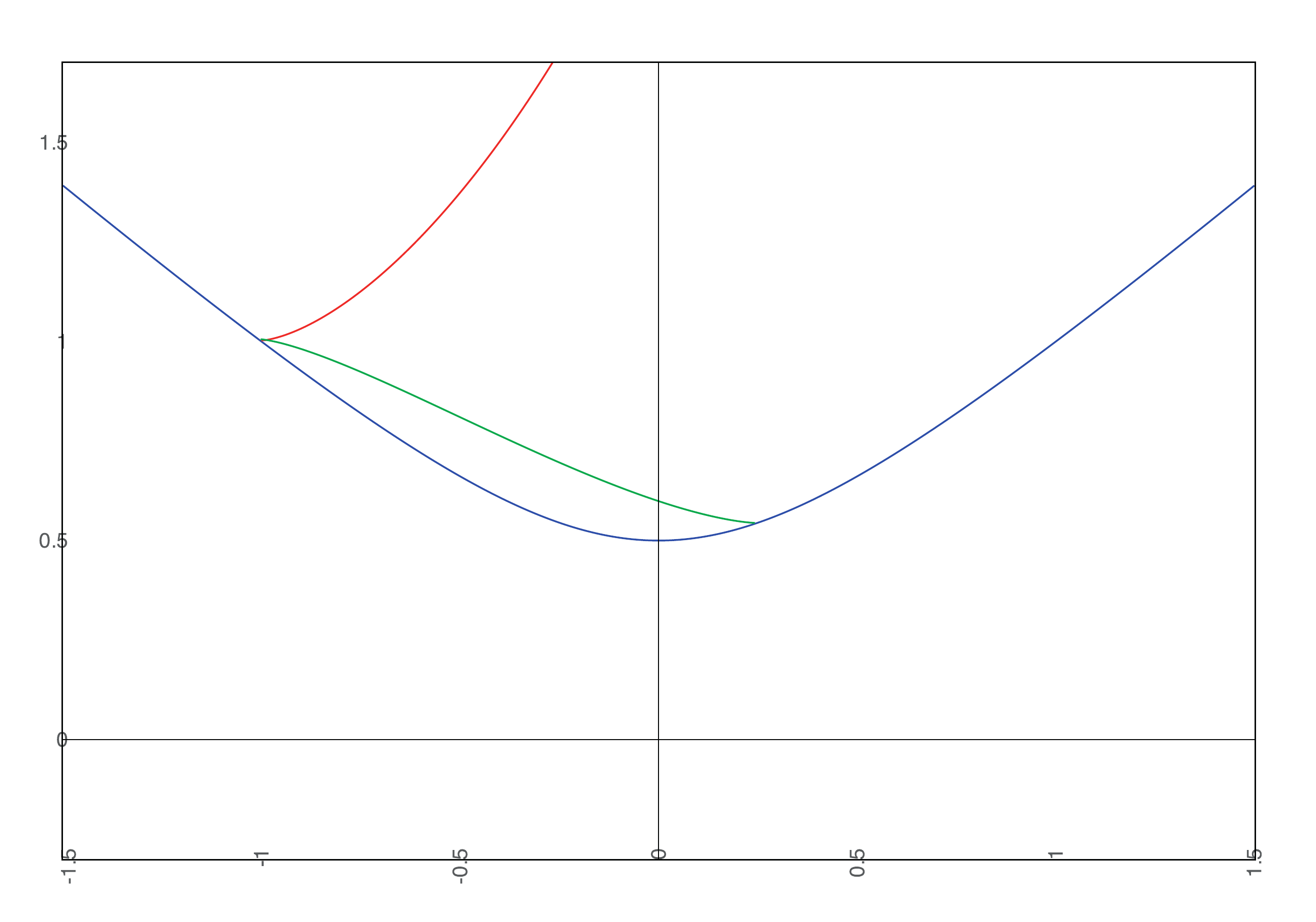},\hs{7} \includegraphics{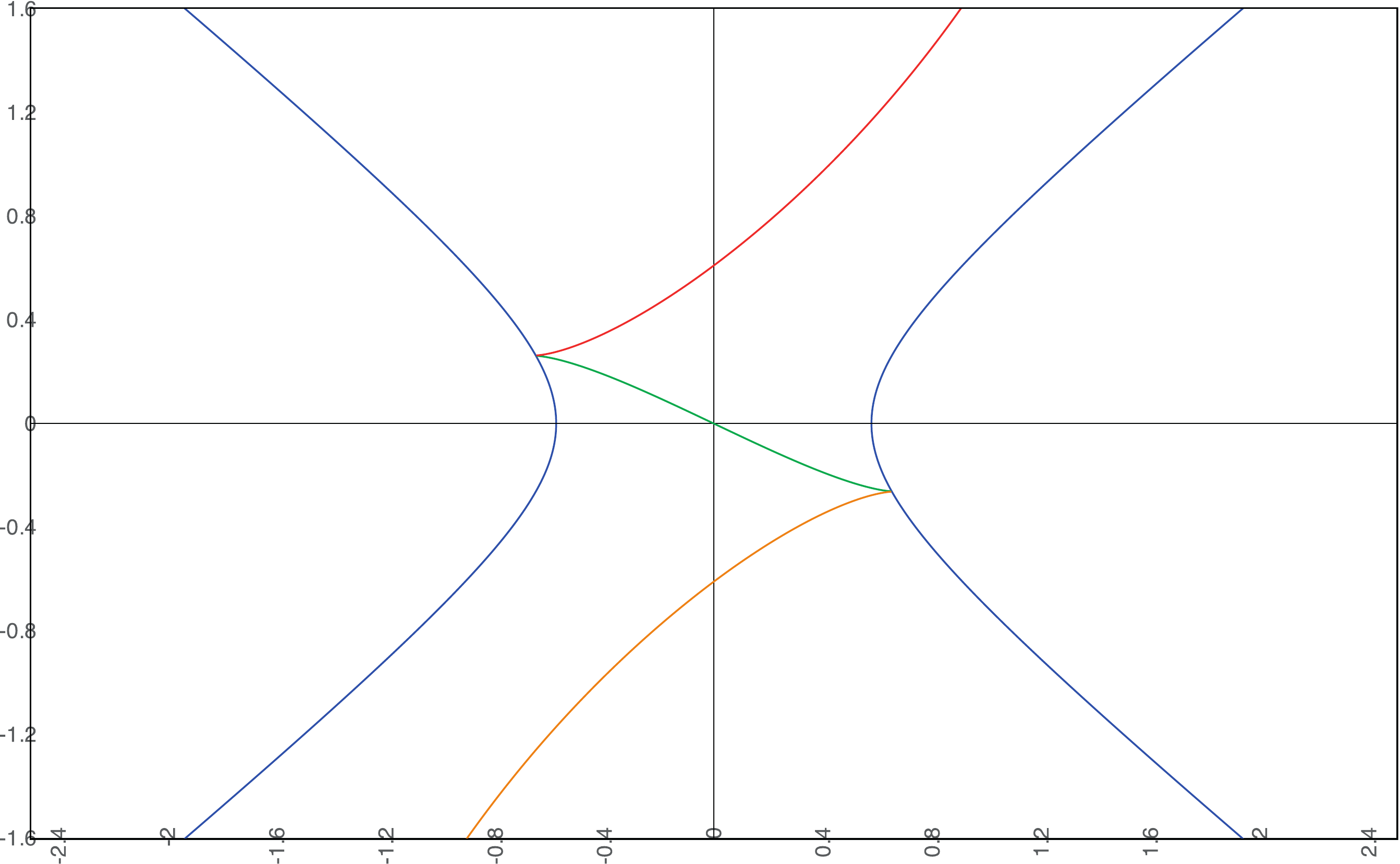}}
\vs{1}

{\footnotesize{\hs{1} Fig.\ 3: Solutions $H[\vf]$ for quadratic potentials (\ref{3.3}) with $v_0 > 0$ (left), 
 and $v_0 < 0$ (right).}}
\ec

\nit
Substitution into eq.\ (\ref{1.5}) then leads to the equalities
\be
3 h_0^2 - 2 h_1^2 = v_0, \hs{1} h_1(3h_0 - 4 h_2) = 0, \hs{1} 4h_1(h_1 - 4 h_3) 
 + \frac{8}{3}\, h_2 \lh 3 h_0 - 4 h_2 \rh = m^2, \hs{1} ...,
\label{3.5}
\ee
from which the solutions $H[\vf]$ can be reconstructed. The same information can be used to calculate 
the total expansion factor of the universe, as defined by the number of $e$-folds in some interval of time:
\be
N = \int_1^2 dt H = - \int_1^2 d\vf\, \frac{H}{2H'} = - \frac{1}{2}\, \int_1^2 d\vf\, 
 \frac{h_0 + h_1 \vf + h_2 \vf^2 + ...}{h_1 + 2 h_2 \vf + 3 h_3 \vf^2 + ...}.
\label{3.6}
\ee
This number can get sizeable contributions only in regions where the slow-roll condition is satisfied:
\be
\eps = - \frac{\dot{H}}{H^2} = \frac{2H^{\prime\, 2}}{H^2} < 1 \hs{1} \Rightarrow \hs{1} 
3H^2 - V < H^2.
\label{3.7}
\ee
Thus we simultaneously have
\be
V < 3 H^2 \hs{2} \mbox{and} \hs{1} V > 2H^2 \hs{1} \Leftrightarrow \hs{1} 
0 \leq \frac{V}{3} < H^2 < \frac{V}{2}.
\label{3.8}
\ee
Not surprisingly, this is the regime in which the kinetic energy of the scalar field is relatively small. 
In most cases it holds only for a relatively narrow range of field values, keeping the number of
$e$-folds near a turning point small. Similar constraints derived from dilaton dynamics 
severely restrict inflationary scenarios in string cosmology \ct{brustein-steinhardt1993}.  

\section{Unstable endpoints and inflation \label{s4}} 

In the above discussions we have carefully distinguished between end points and turning points of the 
scalar field evolution. In both cases $\dot{\vf} = 0$, but at end points in addition $\ddot{\vf} = 0$, which 
can happen only at extrema of the potential $V[\vf]$. However, if the end point occurs at a relative 
maximum or saddle point of the potential, this end point will be classically unstable. Indeed, the field 
can remain there for an indefinite period of time, but any slight change in the initial conditions will 
cause it to move on to lower values of the Hubble parameter. Nevertheless, such a period of 
temporary slow roll of the field creates the right conditions for a period of inflation. We finish this 
paper by discussing an example of this kind taken from ref.\ \ct{vholten2002}. 
\bc
\scalebox{0.22}{\includegraphics{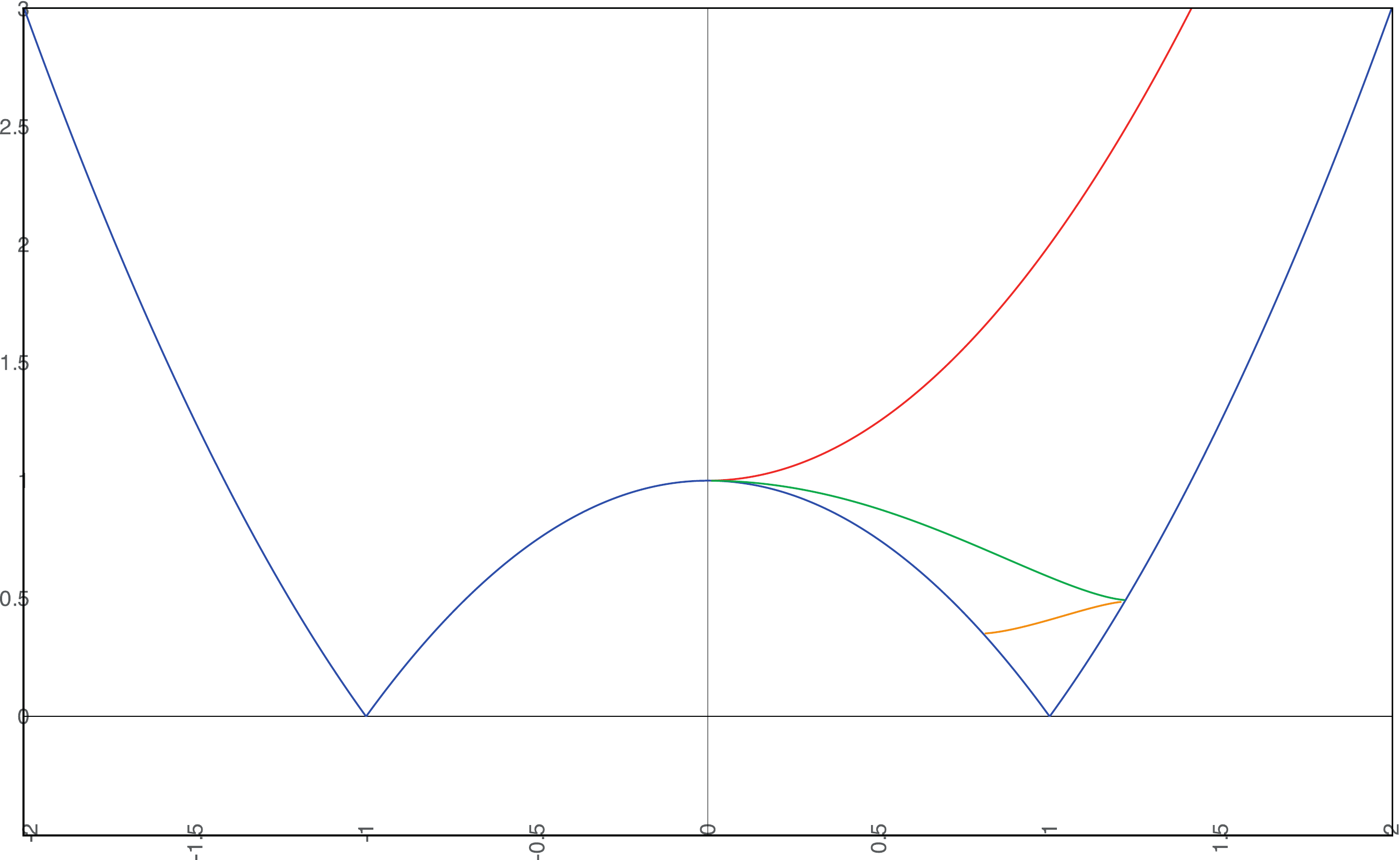}}
\vs{1}

{\footnotesize{Fig.\ 4: Critical curves of stationary points and solutions $H[\vf]$ for a quartic potential with 
spontaneous symmetry breaking. }}
\ec

\nit
This example is tailored to provide an exponentially decaying scalar field 
\be
\vf(t) = \vf_0 e^{-\og t}.
\label{s4.1}
\ee
The Hubble parameter and potential giving rise to this solution can be constructed following the same procedure as
for the eternally oscillating field in sect.\ \ref{s2}, with the result 
\be
H = h + \frac{1}{4}\, \og \vf^2, \hs{2} V[\vf] = v_0 - \frac{\mu^2}{2}\, \vf^2 + \frac{\lb}{4}\, \vf^4,
\label{s4.2}
\ee
where
\be
v_0 = 3 h^2, \hs{2} 
\mu^2 = \og^2 - 3 \og h, \hs{2} \lb = \frac{3 \og^2}{4}.
\label{s4.3}
\ee
Thus we obtain a quartic potential; for $\mu^2 > 0$ it has minima in which reflection symmetry is spontaneously 
broken. The exponential solution (\ref{s4.1}) ends asymptotically at the unstable maximum of the potential where 
$\dot{\vf} = \ddot{\vf} = 0$. As such it represents an end point of the evolution, but a minimal change in the 
initial conditions for the scalar field will turn the end point into a reflection point (if it starts at lower $H$), or 
it will overshoot the maximum (if it stars at higher $H$). Thus the end point is unstable, but the exponential 
decay (\ref{s4.1}) will still provide a good approximation to first part of the evolution of the universe for 
solutions $H[\vf]$ coming close to the maximum of the potential. 

Next observe, that the Hubble parameter (\ref{s4.2}) in combination with the exponential scalar field (\ref{s4.1}) 
leads to a behaviour of the scale factor given by
\be
a(t) = a_0\, e^{ht + \frac{1}{8} \vf^2_0\, (1 - e^{-2\og t})}. 
\label{s4.4}
\ee
Thus for $h > 0$ this epoch in the evolution of the universe ends in an asymptotic de Sitter state with 
Hubble constant $h$. Afterwards, the scalar field will roll further down the potential; provided $3h \leq \og \leq 6h$ 
it will oscillate around the minimum until it comes to rest in another de Sitter or a Minkowski state, again depending 
on the value of $h$. In particular, for $\og \geq 3h$ the model has a final de Sitter or Minkowski state in which
$\dot{\vf} = 0$ and
\be
\langle \vf^2 \rangle = \frac{\mu^2}{\lb} = \frac{4}{3} \lh 1 - \frac{3h}{\og} \rh.
\label{v2.1}
\ee
In this final state the energy density is
\be
\langle V \rangle = v_0 - \frac{\mu^4}{4\lb} = \frac{\og}{3} \lh 6h - \og \rh.
\label{v2.2}
\ee
On the other hand, the energy density for the solution (\ref{s4.1}) is
\be
\rg_s(t) = \frac{1}{2}\, \dot{\vf}^2 + V = 3 H^2 = 3 \lh h + \frac{1}{4}\, \og\, \vf_0^2\, e^{-2 \og t} \rh^2.
\label{v2.3}
\ee
Now the solution (\ref{s4.4}) for the scale factor shows, that before reaching the first turning point
at $\vf = 0$ the scale factor increases by an additional number of $e$-folds given by
\be
N = \frac{1}{8}\, \vf_0^2.
\label{v2.5}
\ee
Therefore the initial energy density at $t = 0$ can be written as
\be
\rg_s(0) = 3 \lh h + 2N \og \rh^2.
\label{v2.6}
\ee
If we take this initial energy density to equal the Planck density: $\rg_s(0) = 1$, this establishes a 
simple relation between $h$, $\og$ and $N$.

Another relation is obtained by taking $\langle V \rangle$ equal to the observed energy density of
the universe today  \ct{planck2013}:
\be
\langle V \rangle = 3 H_0^2 = 1.04 \times 10^{-120}
\label{v2.7}
\ee
in Planck units. This implies that to an extremely good approximation $\og = 6h$, and 
\be
3 h^2 \lh 1 + 12 N \rh^2 = 1, \hs{2} \mu^2 = 18 h^2, \hs{2} \lb = 27 h^2.
\label{v2.8}
\ee
The lower limit on $N$ for inflation as derived from the CMB observations is $N \geq 60$, implying 
\be
h \leq 0.8 \times 10^{-3}.
\label{v2.9}
\ee
Now expanding $\vf$ around its vacuum expectation value
\be
\vf = \frac{\mu}{\sqrt{\lb}} + \chi,
\label{v2.10}
\ee
the quadratic term in the potential becomes
\be
\Del V = \frac{1}{2}\, m^2_{\chi} \chi^2, \hs{2} m_{\chi} = 6 h \leq 0.48 \times 10^{-2}.
\label{v2.11}
\ee
Converting to particle physics units, this is equivalent to an upper limit on the mass of 
$m_{\chi} \leq 1.2 \times 10^{16}$ GeV. This suggests that the inflaton could be a GUT scalar of 
Brout-Englert-Higgs type \ct{vholten2002}. 
\vs{3}

\section{Discussion \label{s5}}

The homogeneity and isotropy of the large-scale universe and its accelerating expansion suggest 
cosmological scenarios in which scalar fields are an important ingredient of the dynamics. The 
results of the Planck mission seem to agree with single-scalar field scenarios for an early epoch 
of inflation \ct{planck2013}. On the other hand, the accelerating expansion of the universe in 
recent times is often attributed to quintessence, which can also be described effectively by a 
scalar field \ct{wetterich1988,zlatev_etal1999}. 

For these reasons an analysis of scalar-field driven cosmology both at early and at late times is
relevant for understanding the observed cosmic expansion. In this paper we have derived and 
described some quite general features of single-scalar field scenarios. For example, end-points 
of scalar field evolution can occur at non-negative extrema of the potential only, whilst regions 
of negative scalar-field potential generically lead to an eternally evolving scalar field and recollapse 
of the universe. These general features were illustrated by specific examples, mostly chosen for 
simplicity though not necessarily for realistic phenomenology. 

Slow-roll conditions, required to realize inflation, are satisfied not only at extrema of the potential, 
but also near turning points where the kinetic energy of the scalar field vanishes. As expected, in 
general the number of $e$-folds of inflation near any single turning point is small, unless it occurs in 
a region where the potential is rather flat. This happens in particular near a maximum or inflection 
point. We have constructed a Higgs-type of potential which can generate at least 60 $e$-folds of 
inflation for fields with masses in the GUT domain. Interestingly the requirement of a sufficient 
number of $e$-folds imposes an upper bound on the late-time expansion rate represented by 
$h$, as manifest in eq.\ (\ref{v2.8}). Of course, in the case at hand the bound is not very 
restrictive as it exceeds the observed value by many orders of magnitude. 
\vs{3}

\nit
{\bf Acknowledgement} \\
The work described in this paper was presented at the 25th meeting on Physics Beyond the Standard Model 
(Bad Honnef, Germany; March 20, 2013). It  is part of the research program of the Foundation for Fundamental 
Research of Matter (FOM).

\np

\end{document}